\documentclass{JHEP3}
\usepackage{amsmath,amsfonts,yfonts,}
\usepackage{epstopdf}
\usepackage{epsfig}
\allowdisplaybreaks[2]

\newcommand{\be}{\begin{equation}}
\newcommand{\ee}{\end{equation}}
\newcommand{\bea}{\begin{eqnarray}}
\newcommand{\eea}{\end{eqnarray}}
\newcommand{\ba}{\begin{eqnarray}}
\newcommand{\ea}{\end{eqnarray}}
\newcommand{\nn}{\nonumber \\}

\newcommand{\beq}{\begin{equation}}
\newcommand{\eeq}{\end{equation}}
\newcommand{\beqa}{\begin{eqnarray}}
\newcommand{\eeqa}{\end{eqnarray}}
\newcommand{\beqar}{\begin{eqnarray*}}
\newcommand{\eeqar}{\end{eqnarray*}}
\newcommand{\eg}{{\it e.g.,}\ }
\newcommand{\ie}{{\it i.e.,}\ }

\newcommand{\rar}{\rightarrow}

\title{Charged black holes in Ho\v rava gravity}

\author{Stefan Janiszewski, Andreas Karch, Brandon Robinson, David Sommer\\
Department of Physics, University of Washington, Seattle, Wa, 98195-1560, USA\\
Email: {\email{stefanjj@uw.edu}, \email{akarch@uw.edu}, \email{robinb22@uw.edu}, \email{dsommer2@uw.edu}}}

\date{\today}
\abstract{We explore static spherically symmetric black hole solutions allowing a bulk $U(1)$ vector field in the khronometric formulation of Ho\v rava gravity by way of Einstein-\AE ther.  We examine analytic solutions and study numerical results in the limit that the khronon does not backreact on the metric.}

\preprint{}
\begin{document}
\section{Introduction}

There has recently been increased interest in a particular theory of gravity that manifestly breaks Lorentz invariance and is power counting renormalizable: Ho\v rava gravity \cite{Horv}. One of the attractive features of Ho\v{r}ava gravity is that even in the absence of matter it allows solutions with anisotropic scaling between space and time \ie $t\rar \lambda^z t$ and $x_i \rar \lambda x_i$, where $z$ is the dynamical critical exponent, even in the low energy limit where only two-derivative terms are kept in the action. The degrees of freedom are the familiar quantities of the ADM formulation of GR: the lapse $N$, shift vector $N_I$, and spatial metric $G_{IJ}$. The realization of solutions with Lifshitz scaling provide another possible utilization of holography that capture features common to condensed matter systems \cite{AKSJ}.  

One of the most interesting questions in Ho\v{r}ava gravity, as well as similar non-relativistic constructions as summarized in \cite{BlasPS}, is to what extent horizons and their thermodynamic properties, which are such a paradigmatic feature of relativistic gravity, continue to exist in the non-relativistic setting. Our main tool to address these questions is the equivalence between the low energy limit of Ho\v rava and another well studied theory of gravity endowed with a preferred notion of time, Einstein-\AE ther (E-\AE).  The nature of the equivalence was argued in \cite{Berg} to arise in static, spherically symmetric geometries.  It was also recently conjectured in \cite{Jac} that the correspondence between Ho\v rava gravity and E-$\AE$ runs deeper.  The author of \cite{Jac} argues that in the limit that the coupling of the `twist' of the \ae ther vector goes to infinity E-$\AE$ descends to the khronometric formulation of Ho\v rava gravity, and one can also find solutions of the khronometric theory that are not hypersurface orthogonal Einstein-\AE ther\, solutions \eg slowly rotating black holes.  Here, however, we will focus on static, spherically symmetric geometries equipped with a preferred notion of time that allows a geometric construction of both theories based on leaves of constant global time foliating the space-time \cite{BlasPS}.  Transformations that preserve the global time coordinate are good symmetries of the two theories.  In Ho\v rava and Einstein-\AE ther, we have foliation preserving diffeomorphisms (Fdiffs) that include time reparameterizations $t \rar \tilde{t}(t)$ and spatial diffeomorphisms $x_I \rar\tilde{x}_I\left(x\right)$.  In general in Ho\v{r}ava gravity, time dependent spatial diffeomorphisms are allowed \cite{BlasPS}.

Importantly, the Ho\v rava and E-$\AE$ duality affords a window into interesting solutions realizing causal horizons in a natural way.  Since Lorentz invariance is broken, there can be modes propagating faster than the speed of light $c$, which means that light cones are no longer the objects encoding causality.  Instead, causality is enforced by the requirement that propagation is unidirectional into the future according to the preferred time. Recent work in \cite{Blas} has shown that the foliation naturally captures causal horizons as the boundary surface at timelike infinity.  This surface is called the ``universal horizon'' as it represents a trapping surface for arbitrarily fast propagating modes.

This leads one naturally to ask how universal these horizons are; do they appear outside of the asymptotically flat backgrounds; do they persist with the inclusion of parameters beyond mass characterizing the solution \eg charge, angular momentum?  The work done in \cite{SJ} has shown that universal horizons do exist in asymptotically Anti-de Sitter (AdS) spacetimes.  The realization of universal horizons in the probe limit lead to the search for analytic solutions realizing the same behavior.  Recently, such solutions were constructed systematically in \cite{SJ}.

The primary purpose of this paper is to explore more general geometrical settings.  Can we successfully identify a causal boundary in the probe limit with the inclusion of electric charge on the horizon?  Can we then turn on finite couplings and find black hole solutions in the fully backreacting theory?  Can we determine consistently the thermodynamics of such solutions?  Is there a way to realize a holographic picture on the resulting background, and can we interpret the results using language familiar to us for relativistic backgrounds?

In section 2, we will discuss the relationship between spherically symmetric Einstein-\AE ther$\,$and the low energy Ho\v rava theories.  In addition, we will review the results obtained for the asymptotically flat and AdS cases.  In section 3, we will test the universality of the appearance of the causal boundary in an AdS-Reissner-N\"ordstrom background and discuss the subtleties as the charge is increased toward extremal values.  We will then obtain an analytic solution for charged backgrounds, attempting both asymptotically AdS and flat geometries, and examine the resulting near-horizon structure.  Lastly in section 4, we will compute the two point correlation of non-relativistic charged scalars in the near-horizon geometry (\eg $AdS_2 \times R^2$) and attempt an interpretation. In particular, we will compare those results to those of the relativistic charged scalars using the $AdS_2/CFT_1$ correspondence. 

\section{Khronons and universal horizons}

Here we offer a brief overview of the relevant concepts that will be useful for the following sections.  For a comprehensive look at the khronon formalism and its uses see \cite{AKSJ,Berg,SJ,HMT}.  As mentioned above, we can understand the low energy effective Ho\v rava theory in a purely geometric way.  The degrees of freedom available to us in a theory with Fdiff invariance come in the form of the the lapse $N$, shift vector $N_I$, and spatial metric $G_{IJ}$ familiar in the ADM decomposition of the metric in standard GR,
\beq
ds^2 =-N^2dt^2+G_{IJ}\left(dx^I +N^I dt\right)\left(dx^J + N^J dt\right).
\eeq
The low energy Ho\v rava action is
\beq\label{eq:IHor}
I_H = \frac{1}{16\pi G_H}\int \mathrm{d}t\mathrm{d}^3x N\sqrt{G} \left( K_{IJ}K^{IJ}-(1+\tilde{\lambda})K^2+(1+\tilde{\beta})(R-2\Lambda)+\tilde{\alpha}\frac{\nabla_I N\nabla^I N}{N^2}\right),
\eeq
where $K_{I J}=\frac{1}{2N}\left(\partial_t G_{IJ}-\nabla_I N_J -\nabla_J N_I\right)$ is the extrinsic curvature of the spatial leaves foliating the spacetime, $G$ is the determinant of the spatial metric, and $\Lambda$ is the cosmological constant.  The couplings $\left(\tilde{\alpha},\tilde{\beta},\tilde{\lambda}\right)$ enter due to the Fdiff symmetry, and $G_H$ sets the Planck mass.

Fdiffs allow for time reparameterizations, $t\rar \tilde{t}(t)$, which lets us rewrite the low energy Ho\v rava gravity action in a covariant form by way of a scalar field, $\phi(t)$ called the khronon, coupled to GR.  This gives us a point of contact between Ho\v rava and E-$\AE$ wherein the \ae ther vector, $u^M$ can be written in terms of the khronon:
\beq
u_M =\frac{-\partial_M \phi}{\sqrt{-g^{PQ}\partial_P\phi\partial_Q\phi}}.
\eeq
$u_M$ is a time-like unit vector normal to the spatial leaves of the foliation.  By making this identification, and an appropriate rescaling and renaming of couplings, ~\eqref{eq:IHor} can be recast as
\beq\label{Ikh}
I_{kh}=\frac{1}{16\pi G_{kh}}\int\mathrm{d}^4x\sqrt{-g}\left(R-2\Lambda+c_4 u^M\nabla_M u^N u^P\nabla_P u_N-c_2\left(\nabla_M u^M\right)^2-c_3\nabla_M u^N \nabla_Nu^M\right)
\eeq
with the following relations
\beq
\frac{G_H}{G_{kh}}=1+\tilde{\beta}=\frac{1}{1-c_3},\qquad \tilde{\alpha}=\frac{c_4}{1-c_3},\qquad 1+\tilde{\lambda}=\frac{1+c_2}{1+c_3}.
\eeq
After finding the desired analytic solutions, we will need to map to the adapted co\"ordinates of the ADM form of the metric to find the Ho\v rava degrees of freedom.  

The speed for the scalar mode, $s_0$, and the spin-2 graviton $s_2$ can be determined from the linearized theory around a flat background in the weak field, low speed limit:
\beq
s_2^2=\frac{1}{1-c_3},\qquad s_0^2=\frac{\left(c_2+c_3\right)\left(D-1+c_4\right)}{c_4\left(1-c_3\right)\left(D-1+Dc_2+c_3\right)}.
\eeq
In the probe limit, which we will be considering in the next section, the couplings are taken to small values $c_i <<1$, and the speed of the two present modes become
\beq
s_2^2 \approx 1, \qquad s_0^2 \approx \frac{c_2+c_3}{c_4}.
\eeq

\subsection{Asymptotically hyperbolic results}\label{sec: aAdS}

The work in \cite{SJ} is instructive to understand as it provides a basis for further exploration in charged backgrounds.  The primary goal in \cite{SJ} was to realize black hole solutions to \eqref{Ikh} that had an asymptotic geometry amenable to holographic descriptions \cite{AKSJ}.  That is, far from the interior of the $d+1$ dimensional space-time ($\ie$ as $r\rar 0$) the metric has the form
\beq\label{eq:hyp}
ds^2 \approx -\left(\frac{L^2}{r^2}\right)^{z}dt^2 +\frac{L^2}{r^2}\left(dr^2 +d\vec{x}^2\right),
\eeq
where $d\vec{x}^2$ is a section on the $d-1$ dimensional plane.    For the rest of the paper we will be using $d=3$.

The translational invariance on the spatial section indicates that we will be looking for planar horizon solutions (black branes).  Working in the Poincar\'e-like co\"ordinates above lends itself to numerical analysis near the boundary, which will prove useful in the probe limit calculations.  Further, it is more natural to use a co\"ordinate system that covers the whole manifold of a black hole space-time, in-falling Eddington-Finkelstein co\"ordinates (IEF).  This choice will again give a more natural setting in which to perform the probe calculations.  In IEF, we define the in-falling time $v=t-r^*$ where the tortoise co\"ordinate $r^*=\int \mathrm{d}r \sqrt{-\frac{g_{rr}}{g_{tt}}}=\frac{L}{z} (\frac{r}{L})^z$ such that, setting $L=1$, the new metric is
\beq
ds^2=-\frac{dv^2}{r^{2z}}-\frac{2}{r^{z+1}}dvdr+\frac{d\vec{x}^{\,2}}{r^2}.
\eeq
This provides us with the asymptotic form of the generic ansatz for a spherically symmetric black hole space-time,
\beq\label{metsph}
ds^2=-e(r)dv^2-2f(r)dvdr+\frac{d\vec{x}^{\,2}}{r^2}.
\eeq
Now that we have in-hand an ansatz for the metric, we need to take care of the \ae ther vector, which asymptotically should reproduce Poincar\'e time ($\phi \sim t$) of eq.~\eqref{eq:hyp}. In IEF this requires,
\beq
u_M = \left(-\frac{1}{r^z},-\frac{1}{r},0,0\right),
\eeq
for the generic timelike unit vector respecting the same symmetries
\beq\label{aether}
u_M=\left(-\frac{a(r)^2e(r)+f(r)^2}{2a(r)f(r)},-a(r),0,0\right).
\eeq
Since we are respecting spherical symmetry in $d=3$, we can note that hypersurface orthogonality demands the curl of $u_M$ given by $\omega_M = \epsilon_{MNPQ}u^N\nabla^P u^Q$ vanishes.  We can then rewrite the $c_4$ term in eq.~\eqref{Ikh}  as
\beq
u^M\nabla_M u^N u^P\nabla_P u_N -\omega_M\omega^M = -\frac{1}{2}F_{MN}F^{MN},
\eeq
with $F_{MN}=\partial_M u_N -\partial_N U_M$ playing the role of the `field strength'.  The zeroth order solution to the equations of motion, found by varying with respect to $e(r)$, $f(r)$, and $a(r)$ and expanding each as their asymptotic forms multiplied by a power series in $r$, necessitates $c_4=\frac{z-1}{z}$ and, in arbitrary $(d+1)$ dimensions, $\Lambda = -\frac{1}{2}\left(z+d-2\right)\left(z+d-1\right)$.  For example, the asymptotically $AdS_4$ solution is given by $z=1$, $c_4 = 0$, and $\Lambda = -3$.

In making the asymptotic expansions of the metric functions and the khronon, we can solve for the coefficients in the expansions order by order in the hope to obtain an analytic expression for $e(r)$ and $f(r)$.  Specifically, we would hope that the expansions for the metric functions terminate at some order and leave us with a small number of constants describing the solution.   At the very least, we would desire an expansion that has an obvious analytic structure.  Indeed, it was found that for $z=1$, the expansions for both $e(r)$ and $f(r)$ terminate quickly, at least in checking to $\mathcal{O}(r^{30})$, while $a(r)$ does not but is controlled by two parameters $C_e$ and $C_a$.  The appearance of two parameters is troubling at first, but when one imposes regularity at the horizon for the scalar mode, one parameter is immediately fixed.  Once the truncated functions are in hand, we will make the ansatz $a(r)=-\frac{1}{r}b(r)$, where the asymptotic behavior $-\frac{1}{r}$ is multiplying an analytic function $b(r)$.  This assumption of the analyticity of $b(r)$ does not always produce the correct results in the asymptotically Lifshitz, $z\neq1$, solutions as discussed in \cite{SJ}.  However for the asymptotically AdS black hole, the metric functions and the khronon can be solved exactly and adapted to the ADM-like co\"ordinates of Ho\v rava gravity to yield
\beq
ds^2 = -\frac{\left(r^3-r_*^3\right)^2}{r^2r_*^6}dt^2+\frac{r_*^6}{r^2\left(r^3-r_*^3\right)^2}\left(dr + \frac{r^3\left(r^3-r_*^3\right)}{\sqrt{1-c_3}r_*^6}dt\right)^2+\frac{d\vec{x}^2}{r^2}
\eeq

\subsection{Probe limit}

While the analytic solution found in the previous section is interesting in its own right, the probe limit offers us the opportunity to study Ho\v rava gravity using GR solutions as backgrounds for the khronon.  That happens because in the probe limit the couplings that control the dynamics of the khronon are parametrically small, and so metrics that solve the Einstein equations, solve the gravitational equations of motion for Ho\v rava gravity due to the lack of backreaction. The khronon simply imprints a preferred notion of time on a given solution of Einstein gravity. The result is needing to solve, numerically in most cases, the equations of motion for the scalar khronon on a known background.  To that end, we will quote the method used in \cite{SJ} and the results obtained therein for future use in guiding us with the inclusion of charge.

To begin, we can see that in the limit of parametrically small couplings, we can use as our background for the asymptotically AdS solution the metric for an AdS-Schwarzschild black brane with unit mass,
\beq
ds^2 = \frac{1}{r^2}\left(-\left(1-r^3\right)dv^2 -2dvdr+d\vec{x}^2\right).
\eeq
The \ae ther vector, being a unit timelike vector, in these IEF co\"ordinates then can be taken to have the form (akin to eq.~\eqref{aether})
\beq
u_M = \left(-\frac{1+h(r)}{2r}\sqrt{\frac{1-r^3}{h(r)}},-\sqrt{\frac{h(r)}{r^2(1-r^3)}},0,0\right).
\eeq
The demand that asymptotically the khronon encodes Poincar\'e time requires $h(0)=1$ and thus $u_M|_{r\rar 0} \rar\left(-\frac{1}{r},-\frac{1}{r},0,0\right)$.  The khronon action is then
\beq\label{khron}
I_{kh} = \frac{-c_4}{16\pi G_K}\int\mathrm{d}v\mathrm{r}\mathrm{d}x^2\sqrt{-g}\left(\frac{1}{2}F_{MN}F^{MN}+s_0^2\left(\nabla_M u^M\right)^2\right).
\eeq
It is straightforward to find the equations of motion for $h(r)$, the result is a non-linear ODE. Numerical integration starts by Taylor expanding around the value of the radial co\"ordinate $r_c$ at which $h(r_c)=\frac{1-s_0}{1+s_0}$ for fixed scalar speed $s_0^2$.  A shooting method can then be implemented to determine exactly which value of $r_c$ corresponds to the correct boundary conditions at $r=0$.  The value $r_c$ labels the location of the sound horizon for the scalar mode \cite{Blas}.  It is clear that when the  speed of the scalar and spin-2 graviton coincide the $r_c$ associated with the scalar mode will match that of the metric horizon, which is the sound horizon of the spin-2 mode.

After finding the location of the sound horizon for a given $s_0^2$, numerical integration can be performed from $r_c$ toward the interior ($r\rar\infty$) to determine the location of any singular surfaces.  For $s_0^2<1$, the sound horizon sits outside of the metric horizon ($r_c<r_h$), and for $s_0^2>1$, the sound horizon is interior to the metric horizon.  One must take care as the position of the metric horizon is a singular point of the equations of motion and so must be bridged by matching a Taylor series both on the interior and exterior plus/minus some infinitesimal distance.  

\section{Charged black holes in khronometric Ho\v rava gravity}	

Once the metric horizon has been traversed, we can continue integrating to the interior in search of the position of the universal horizon $r_*$, which is seen when the Eddington-Finklestein time component of the \ae ther vector, $u_v$, vanishes.  In the language of Ho\v rava gravity, $r_*$ is the location at which the lapse vanishes and the leaves of the foliation `pile up', \eg $h(r) = -1$. This behavior is illustrated quite dramatically in fig.~\ref{fig:ScKh}, showing the Penrose diagram of AdS space foliated by the preferred time of Ho\v rava gravity.  It is apparent that there exists a radial co\"ordinate beyond which there is no causal connection to the exterior space no matter the speed of propagation and, as discussed earlier, is given by the leaf of the foliation at timelike infinity.
	\FIGURE[ht]{
	\includegraphics[width=0.6\textwidth,angle=0]{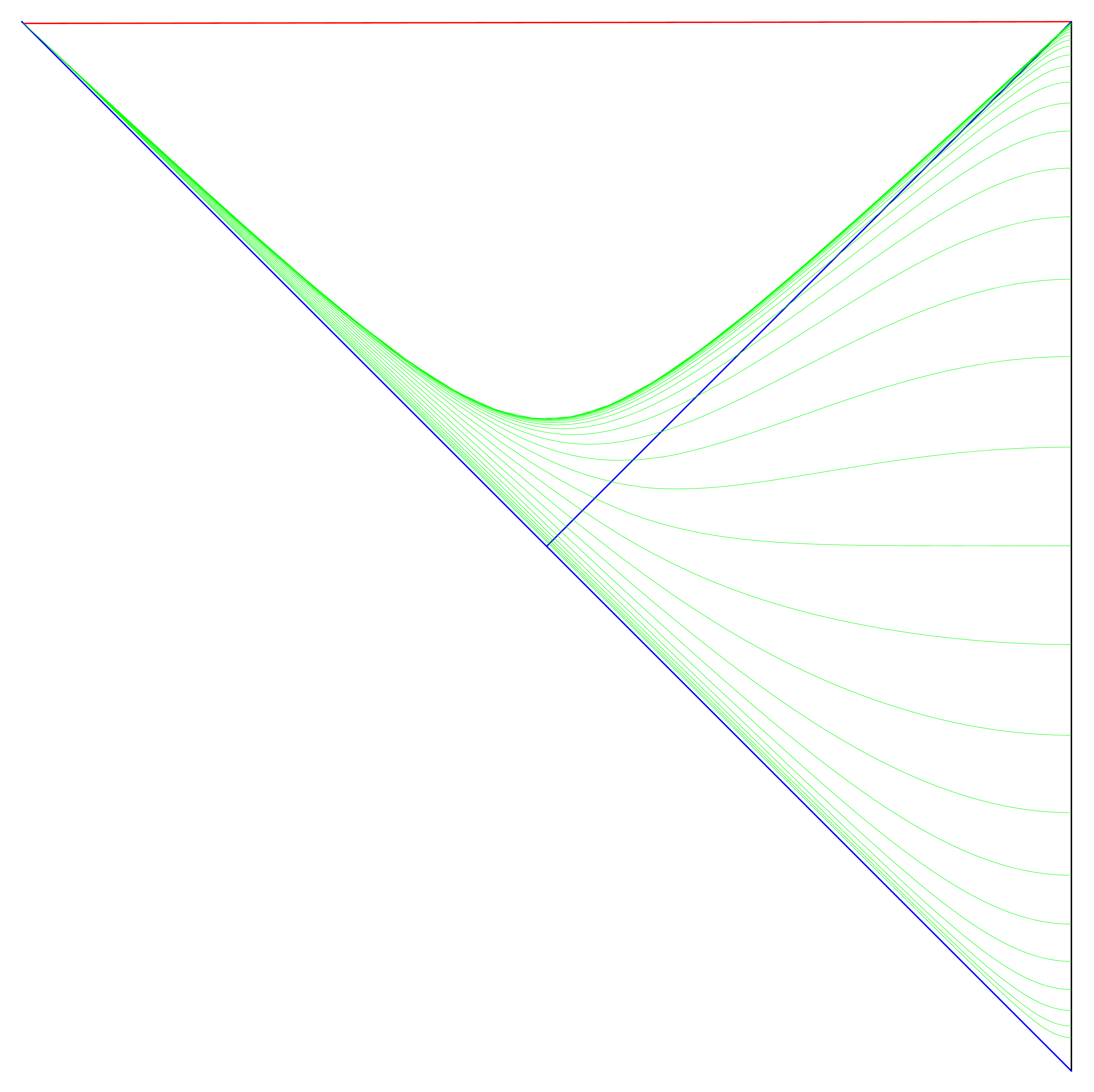}\label{fig:ScKh}
	\caption{Foliation of the AdS-Schwarzschild black hole by leaves of constant $u^\mu$ (green lines).  The vertical black line represents the AdS boundary.  The horizontal red line shows the singularity, and the blue lines at $45^\circ$ represent bounding light rays in the background geometry.}} 
Given the success in finding universal horizons in the probe limit of an AdS-Schwarzschild black brane and asymptotically AdS solutions to the full backreacting theory, we can begin to ratchet up the electric charge.  The hope would be to find that the causal boundaries previously discovered do occur generically in the probe limit.  In addition, we can consider finite values of the couplings and hopefully find interesting backreacted asymptotically AdS charged solutions.

\subsection{Probe limit}\label{sec:probe}
The analysis of the AdS-Schwarzschild black brane carries over to the AdS-Reissner-N\"ordstrom background in the probe limit with the obvious additional parameter of charge, $Q$.  With this additional parameter and knowledge of its problematic limits from literature on charged black holes as GR solutions, it would be wise to fully explore the range of charge from $Q=0$, which will hopefully reproduce results in the previous section, all the way up to extremality where the inner and outer horizons coalesce.  Since we too desire to study the behavior of the khronon on a geodesically complete manifold, it is instructive to consider first the AdS-Reissner-N\"ordstrom metric in the in-falling Eddington-Finklestein co\"ordinates:
\beq
ds^2 = \frac{1}{r^2}\left(-(1-M r^3 + Q r^4)dv^2 - 2 dv dr + d\vec{x}^2\right)
\eeq
where M and Q represent the mass and charge, respectively, of the black brane.

The \ae ther vector in terms of the scalar khronon, $\phi (r)$, for the AdS-Reissner-N\"ordstrom background is:
\beq
u_M=\left(-\frac{1+\phi (r)}{2r\sqrt{\frac{\phi (r)}{1-M r^3 + Q r^4}}},-\sqrt{\frac{\phi (r)}{r^2\left(1- M r^3 + Q r^4\right)}},0,0 \right).
\eeq
	
On this background the probe action for the khronon takes the form,
\beq
I_{khM} = \frac{-1}{16\pi G_K } \int \mathrm{d} t  \mathrm{d}^d x\sqrt{-g}\left(F^{M N}F_{M N} +s_0^2\left(\nabla_N u^{M}\right)^2 +b^2R_{M N}u^M u^N\right)
\eeq
where $F_{M N} = \partial_M u_N - \partial_N u_M$.  The $R_{M N} u^M u^N$ term, while not manifestly zero as in flat backgrounds, does not contribute to the khronon equations of motion.  Varying with respect to $\phi (r)$  we can make a series expansion around the sound horizon $\phi (r_c)=\phi_c$ and solve the equations of motion order by order in $(r-r_c)$ as in the analysis of the AdS-Schwarzschild black brane.

\FIGURE[ht]{
	\includegraphics[width=0.6\textwidth,angle=0]{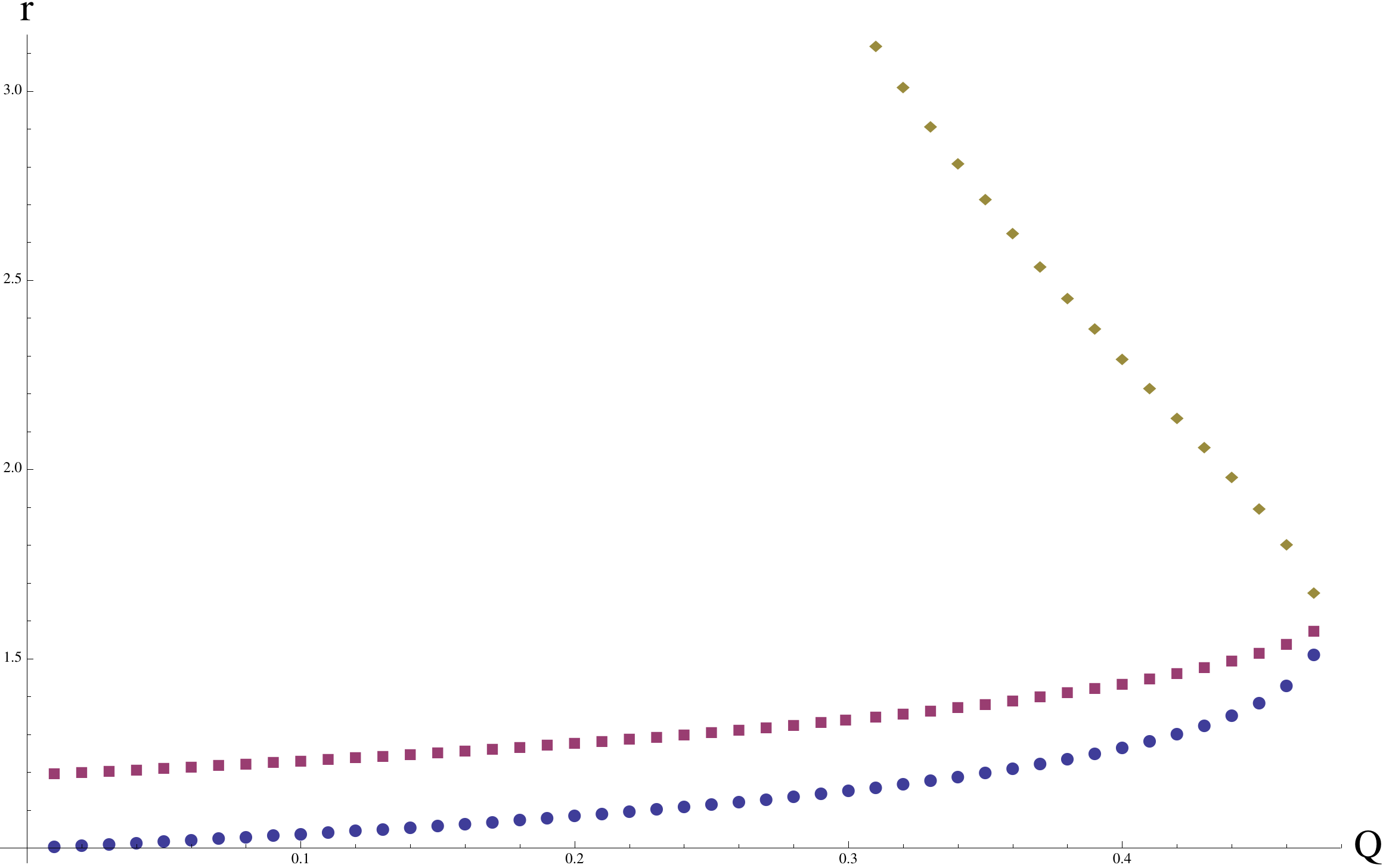}\label{fig:rvQ}
	\caption{Plot of the value of the different horizon radii as we change the charge in the background up to near the critical value $Q_c$. Upper ($\Diamond$) curve represents $r_h^-$, the center ($\Box$) curve shows $r_*$, and the lower ($\bigcirc$) curve displays $r_h^+$.  One can see that the curves coalesce as $Q\rar Q_c$}}
	
Thus, he primary result of the numerical analysis is that as the inner and outer horizons coalesce as $Q\rar Q_c = 3^{1/3}(\frac{3M}{4})^{4/3}$  the universal horizon is being pinched between the two surfaces.  We can see this clearly from fig.~\ref{fig:rvQ} that as we approach criticality, $r_h \approx 1.5863$,  $r_*$ is dragged toward this location.  Upon constructing the Penrose diagrams, fig.~\ref{fig:M1Q25} and fig.~\ref{fig:M1Q4}, as in the AdS-Schwarzschild case, we can see that the leaves of the foliation are piling up nearer to the metric horizon as Q increases.  This observation will play a role in searching for analytic black hole solutions in the next section as we will see this behavior manifest itself in a natural way. 

\DOUBLEFIGURE[t]{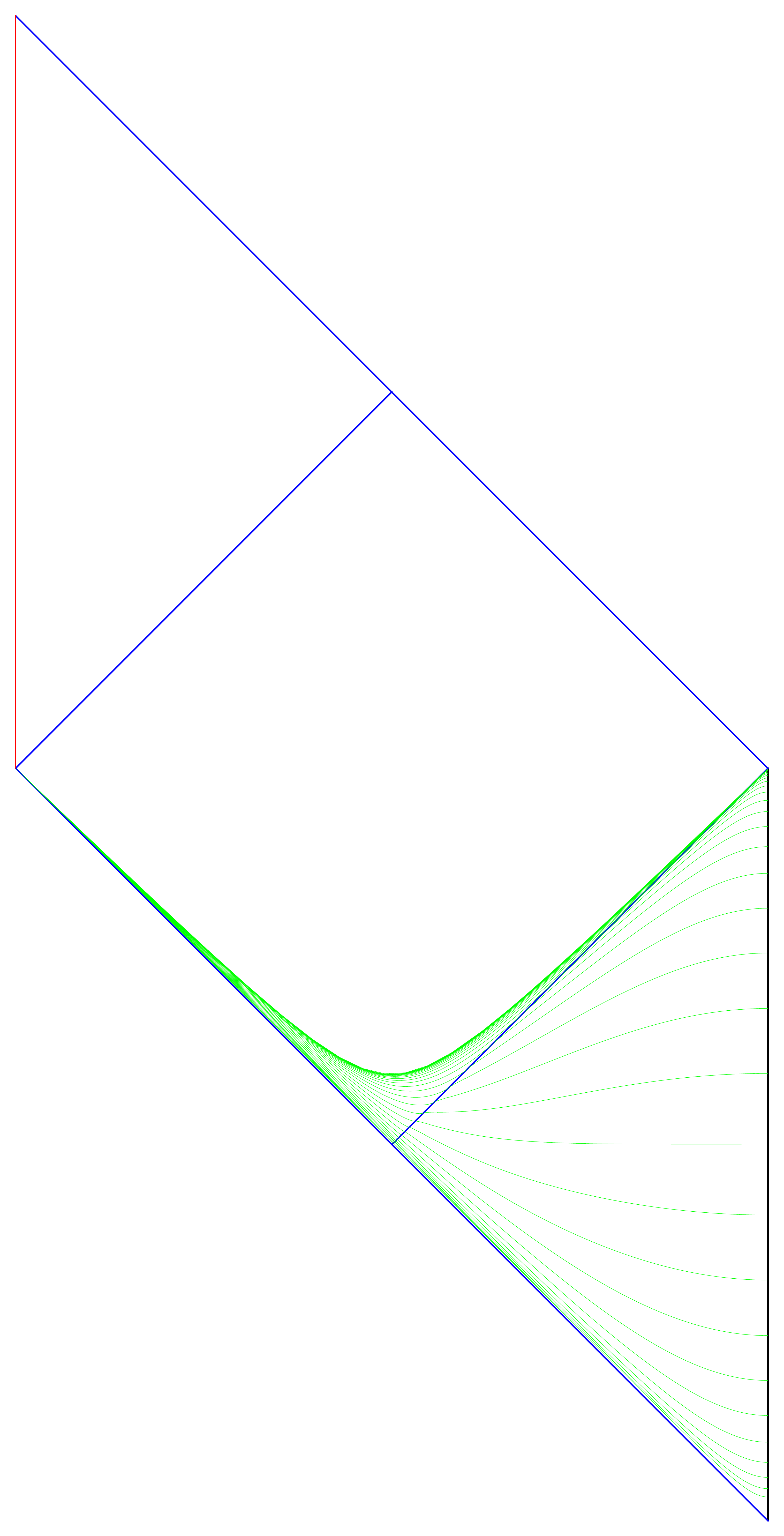, width=0.48\textwidth ,angle=0}
		{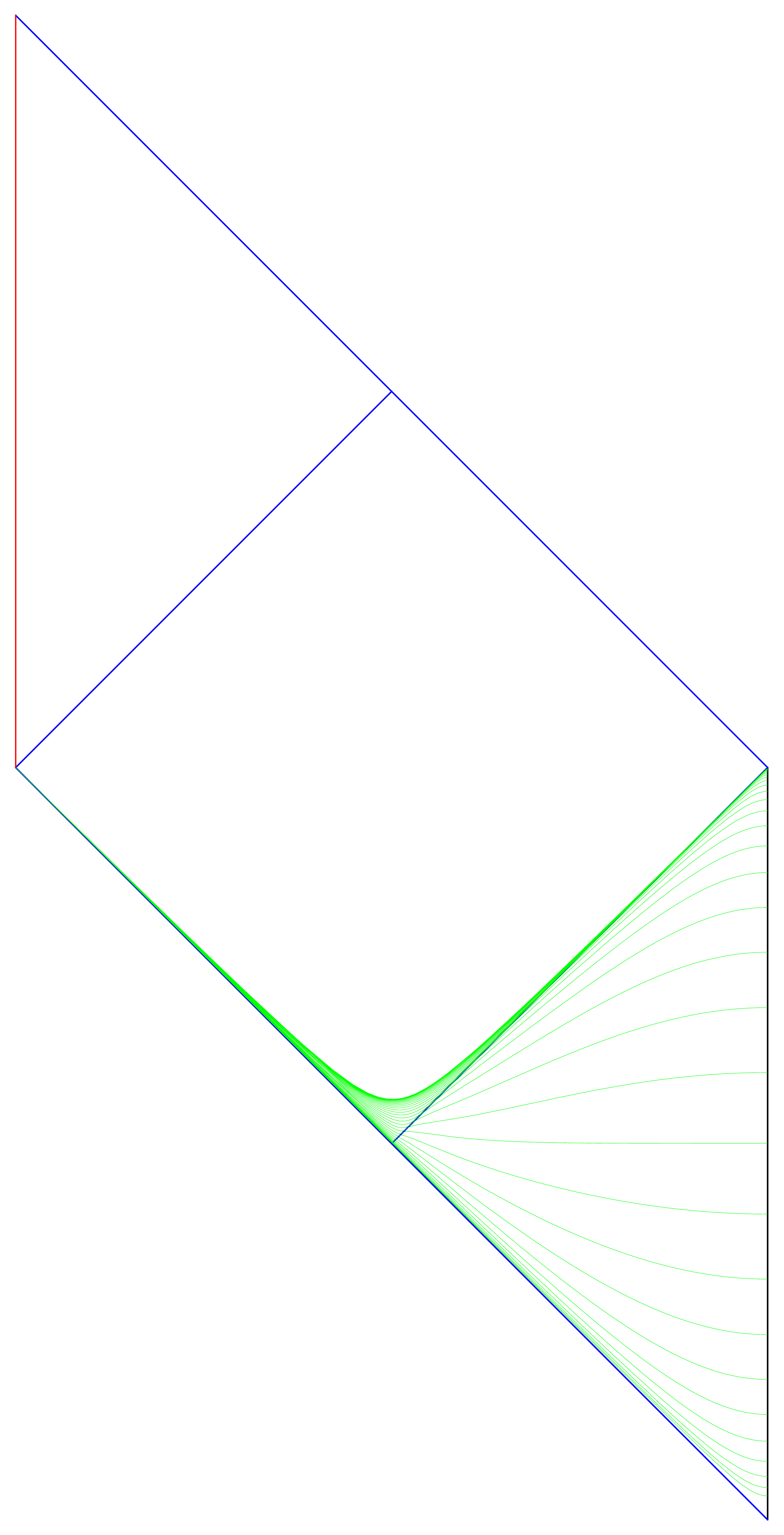, width=0.48\textwidth ,angle=0}
		{Foliation, spatial profile of constant preferred time, of the AdS-RN background with $M=1$ and $Q=0.25$ 			 \label{fig:M1Q25}}
		{Foliation of AdS-RN background with $M=1$ and $Q=0.4$ \label{fig:M1Q4}}
\subsection{Analytic solutions}

Taking the couplings away from the probe limit to finite values, we want to consider the following action for full backreacting khronometric theory
\beqa\label{IKM}
I_{kh} = \frac{1}{16\pi G_K}\int \mathrm{d}t\mathrm{d}^3 x &\sqrt{-g}&\left(R -2\Lambda -\frac{c_4}{2}F_{a b}F^{a b} - c_2 \left(\nabla_b u^{a}\right)^2 -c_3 \nabla_b u^{a} \nabla_a u^{b}\right.\nonumber\\ 
&&\left.\quad+\mu\mathcal{F}_{ab}\mathcal{F}^{ab}+\frac{\kappa}{4}u^a\mathcal{F}_{a b}u^c\mathcal{F}_{c}^{\,\,b}\right)
\eeqa
where $F_{ab}$ is again the khronon ``field strength'', $\mathcal{F}_{a b} = \partial_a \mathcal{A}_b-\partial_b \mathcal{A}_a$ is the electromagnetic field strength, and $\mathcal{A}_t =\rho(r)$ is the gauge potential.  Traditionally, we would only have $\mathcal{F}^2$ in considering the contribution of the Maxwell field, but owing to less restrictive FDiff symmetry, we have the presence of the novel $u^a\mathcal{F}_{a b}u^c\mathcal{F}_{c}^{\,\,b} = E^2$.  When we consider asymptotically flat solutions, $\mathcal{F}^2$ does not play as an important role as it does for asymptotically AdS and will then be disregarded.  Analysis was done to account for the possibility of dyonic black holes with the $\sqrt{-g}\mu\mathcal{F}_{a b}\mathcal{F}^{a b}$ included in \eqref{IKM} and $\mathcal{A}_i \neq 0$ but no fruitful results appeared.

In this section, we will use as our metric ansatz eq.~\eqref{metsph} giving the asymptotic geometry \eqref{eq:hyp} with $z=1$.  We also must specify the asymptotic behavior of $\rho(r)$, which simply amounts to solving Maxwell's equations in $AdS_4$ \ie $\rho(r) \sim Q r$ for small $r$.  With the asymptotic expansions for $e(r)$, $f(r)$, and $a(r)$ from eq.~\eqref{sec: aAdS}, we can solve the equations of motion order by order to $\mathcal{O}(r^{30})$ noting that for $z=1$ the zeroth order requires $\alpha = 0$ and $\Lambda = -3$, which results in
\beqa\nonumber
e(r) &=&\frac{1}{r^2}\left(1-2r^3C_a +\frac{\kappa-8\mu}{24} C_e r^4\right)\\\nonumber
f(r) &=&\frac{1}{r^2}\\\nonumber
a(r) &=&-\frac{1}{r}\left(1 +C_a r^3+\frac{1}{2}\left(3C_a^2-C_e\right)r^6+\frac{1}{2}\left(5C_a^2-3C_e \right)+\ldots \right)\\\nonumber
\rho(r) &=&\sqrt{C_e}r.
\eeqa

We can see that the metric functions and the scalar potential truncate, while the khronon function $a(r)$ does not.  We can appeal to the logic for the uncharged case in using the above expressions as ansatze in our equations of motion with $a(r)=-\frac{1}{r}b(r)$ where $b(r)$ is an analytic function.  Solving the resulting differential equation coming from the $e(r)$ equations of motion for $b(r)$ and fixing the constants by comparing to the $a(r)$ series gives
\beq\label{bsol}
b(r)=\pm\left(1-2C_a r^3+C_e r^4\frac{\kappa-8\mu}{24}\right)^{-\frac{1}{2}}
\eeq
where the $+$ sign is taken for $C_a >0$ and the $-$ sign for $C_a <0$.  Inserting eq.~\eqref{bsol} in the equations of motion, we find satisfaction only when $\kappa=-8\mu$.  However, this solution does not descend smoothly to a probe limit solution, except for the extremal case, and is thus disregarded. Therefore, non-extremal AdS-Reissner-N\"ordstrom is not a physical solution of eq.~\eqref{IKM}, as already seen in the probe limit of section \ref{sec:probe}.

While we only have probe limit solutions with asymptotically AdS  geometry, we can still search for asymptotically flat analytic solutions.  We cannot make a metric ansatz with a planar geometry on the horizon in this case, but instead we will assume the metric has the form
\beq\label{eq:flat}
ds^2 = -e(r) dv^2 -2f(r)dvdr +1/r^2\left(d\theta^2+sin^2\left(\theta\right)d\phi^2 \right),
\eeq
and the \ae ther vector has the form eq.~\eqref{aether}, where asymptotically
\beq
e(r)\sim1,\qquad f(r)\sim\frac{1}{r^2}\qquad a(r)\sim-\frac{1}{r^2}\qquad\rho(r)\sim Q r.
\eeq
Inserting eq.~\eqref{eq:flat} into eq.~\eqref{IKM} and applying the same asymptotic prescription to the resulting equations of motion, we find
\beqa\label{flatbh}
e(r)&=& 1-2C_a r+(C_a r)^2\nn
f(r)&=&\frac{1}{r^2}\nn
a(r)&=&1-C_a r+C_a^2 r^2-C_a^3 r^3+\ldots\nn
\rho(r)&=&2r\sqrt{\frac{(2-\alpha)C_a^2}{\kappa}}.
\eeqa
We can immediately recognize the above solution as the extremal Reissner-N\"ordstrom black hole with the khronon encoding global time.  The latter point comes from formally taking the sum for $a(r)=\frac{1}{1-C_a r} =\sqrt{e(r)^{-1}}$.  One should note that there was no specification of the couplings appearing in eq.~\eqref{IKM}, which is a departure from the uncharged asymptotically AdS solutions.  Further, there is a lone parameter describing our black hole, which is easily seen to be the location of both the metric and universal horizon.  This should be expected from the nature of the behavior of the universal horizon near extremal charge in the probe limit.

In transforming back to the adapted co\"ordinates for Ho\v rava gravity, we find no surprises.  That is, the shift vector $N_I=0$, the lapse $N=\sqrt{-g_{tt}}$, and the spatial metric is just the induced metric on a slice of constant time.  This stands in contrast to the asymptotically AdS solution in section \ref{sec: aAdS} that had a far more interesting structure.  Thus, we have a perfectly good relativistic vacuum as a solution to our non-relativistic theory.

While the full asymptotically flat geometry is not useful in the context of holography, it is well known that for the extremal Reissner-N\"ordstrom the near horizon geometry is that of $AdS_2 \times S^2$.  This is the setting in which we will calculate correlation functions.  It is important then to understand the thermodynamics as we approach the horizon.   So, it is important to understand the thermodynamics of the space-time in both limits.  However, the calculations for the global geometry are trivial as the matter content (\ie the khronon) does not contribute to the calculation of the temperature or free energy, and we simply recover well known results for the extremal Reissner-N\"ordstrom black hole.  There was some hope that the near horizon geometry would remember something about the free energy outside of the throat, but we found what one would expect for a vacuum space-time like $AdS_2 \times S^2$: zero temperature and zero free energy.

As we have found that the near horizon geometry is of particular interest, it is natural to check whether we can find $AdS_2 \times \mathcal{M}^2$ to be a solution of the equations of motion derived from eq.~\eqref{IKM} for $\mathcal{M}^2$ being $R^2$, $S^2$, or $H^2$, that is, a plane, a sphere, or a hyperbolic plane, respectively.  We note that $AdS_2 \times R^d$ was shown to be a solution to Ho\v{r}ava gravity in \cite{AY}.  If we consider the metric ansatz
\beq
ds^2 = -e(r)dt^2 +f(r)dr^2+dx^2 +g(x)^2 dy^2,
\eeq
the desired asymptotic behavior on the $AdS_2$, including the length scale $L$, of the metric functions $e(r)$, $f(r)$, and the background scalar potential $\rho(r)$ is given by 
\beq
e(r)\sim\frac{L^2}{r^2},\quad f(r)\sim\frac{L^2}{r^2},\quad \rho(r)\sim\frac{Q L}{r},
\eeq
with the aether vector
\beq
u_M = \left(-\sqrt{e(r)}\sqrt{\frac{a(r)^2+f(r)}{f(r)}},a(r),0,0\right).
\eeq
Since the asymptotic behavior of the aether vector is to encode AdS time, in the $r\rar 0$ limit $a(r) \sim 0$.  Starting from scratch, we insert the ansatz into the equations of motion, and we find $\Lambda = \frac{-2+c_4+\kappa Q^2}{2L^2}$ is required.  This leaves us to solve for $g(x)$, which yields
\beq
g(x)= c_1e^{\sqrt{1-c_4 -\kappa Q^2}\frac{x}{L}}+c_2 e^{-\sqrt{1-c_4-\kappa Q^2}\frac{x}{L}}.
\eeq
Thus, we see that an $AdS_2 \times R^2$ can be obtained by choosing $Q=\pm\sqrt{\frac{1-c_4}{\kappa}}$ and $\Lambda = -\frac{1}{2L^2}$, resulting in constant $g(x)$.  Furthermore, we can observe that for $Q>\sqrt{\frac{1-c_4}{\kappa}}$
\beq
L_H = \frac{L}{\sqrt{1-c_4-\kappa Q^2}},\quad \Lambda < -\frac{1}{2L^2}, \quad \mathcal{M}^2 = H^2,
\eeq 
where $L_H$ is the hyperboloid curvature.  Additionally, for $Q<\sqrt{\frac{1-c_4}{\kappa}}$,
\beq
L_S = \frac{L}{\sqrt{|1-c_4-\kappa Q^2|}},\quad \Lambda> -\frac{1}{2L^2},\quad \mathcal{M}^2 = S^2
\eeq
with $L_S$ being the $S^2$ curvature.  In comparing to the $\Lambda = 0$ solution found above, we find agreement with the result for $Q$ in eq.~\eqref{flatbh}.  While we may not be able to access the full global black hole solutions that give rise these geometries, we can utilize the near horizon solutions in the next section.

\section{NR holography on the $AdS_2$}
In this section, we explore the holographic description of non-relativistic, charged scalar fields living in the throat of the geometry \eqref{eq:flat}.  That is, given the results of \cite{AKSJ}, we would like to understand the dual non-relativistic $0+1$ dimensional theory to our Ho\v rava solution.  Specifically, we can consider two-point correlation functions on $AdS_2\times R^2$ as shown above.  In doing so, we can compare to the fully relativistic results in \cite{MIT} to probe the differences between employing holography for fields that respect different boundary symmetries on the same background.

Consider the action of a charged scalar with manifest Schr\"odinger symmetry \cite{DSon}
\be
S = \int dt dr d^2 x \sqrt{G} \frac{N^2}{\lambda} \left[ \left( \frac{\imath \lambda}{2 N^2} \Psi^{\dagger} \left( \mathcal{D}_t - N^J \mathcal{D}_J \right)  \Psi + h.c. \right)  - G^{IJ} \mathcal{D}_I \Psi^{\dagger} \mathcal{D}_J \Psi - M^2 \Psi^{\dagger} \Psi \right],
\label{act1}
\ee
where in accordance with the analysis done in \cite{AKSJ}, $\mathcal{D}_t =\partial_t -\imath \mathcal{A}_t$ and $\mathcal{D}_I=\nabla_I-\imath \mathcal{A}_I$ are the gauge covariant derivatives.  The background gauge field is $\mathcal{A}_\mu = \left(\frac{g}{r}, \vec{0}\right)$, $G^{IJ}$ is the inverse spatial metric, $M$ is the mass of the charged scalars, and $\lambda/2$ is their charge.  $N$ is the lapse encoding global time, and the shift vector $N_I$ vanishes in $AdS_2 \times R^2$.

After Fourier transforming in the orthogonal, $(t,\vec{x})$, directions $\Psi=\psi(r) e^{-\imath\omega t}e^{\imath \vec{k}\cdot\vec{x}}$, the equation of motion for the scalar field $\psi$ is
\beq\label{eq:eomz}
\psi''\left(r\right)-\frac{\psi'\left(r\right)}{r}+\left(\lambda\left(\frac{g}{r}+\omega\right)-\frac{\kappa}{r^2}\right)\psi\left(r\right)=0,
\eeq
where $\kappa \equiv k^2 +M^2$.  Equation \eqref{eq:eomz} can be solved exactly in terms of confluent hypergeometric functions of the second kind, $U(a;b;z)$, and generalized Laguerre polynomials $L(a;b;z)$:
\beq\label{eq:hype}
\psi(r) = e^{-i \sqrt{\lambda \omega} r} r^{1+\nu} \left[ \widetilde{C}_1 U\left(\chi; \ 1+2\nu; \ 2 i\sqrt{\lambda \omega} r \right) + \widetilde{C}_2 L\left(-\chi; \ 2\nu; \ 2 i\sqrt{\lambda \omega} r   \right) \right],
\eeq
where $\chi =\frac{1}{2}+\nu + i \frac{g}{2}\sqrt{\frac{\lambda}{\omega}}$ and $\nu \equiv \sqrt{1+\kappa^2}$.
Making an asymptotic expansion of the solution in the bulk ($r \to \infty$), we find that both terms have ingoing and outgoing contributions.  To eliminate the outgoing contributions and thereby reconstruct a retarded Green's function, we can rewrite the solution in terms of Whittaker function of the first kind $\mathcal{M}(a;b;z)$:
\beq\label{eq:whit}
\psi(r) = C_1\sqrt{r}\mathcal{M}\left(-\imath\frac{g}{2}\sqrt{\frac{\lambda}{\omega}};\nu;2\imath r\sqrt{\lambda\omega}\right)+C_2 \sqrt{r}\mathcal{M}\left(-\imath\frac{g}{2}\sqrt{\frac{\lambda}{\omega}};-\nu;2\imath r\sqrt{\lambda\omega}\right),
\eeq
where $C_1$ and $C_2$ are non-trivially related to the original constants $\widetilde{C}_1$, $\widetilde{C}_2$ in \eqref{eq:hype} and generically have $k$ and $\omega$ dependence.  While both terms in \eqref{eq:whit} have ingoing and outgoing contributions, the Whittaker function admits an asymptotic expansion $\mathcal{M}(a;b;z) \sim e^{-z/2} F(a,b,z) + e^{z/2} G(a,b,z)$ that explicitly separates the ingoing and outgoing contributions, and the coefficients $F(a,b,z)$ and $G(a,b,z)$ are given in closed form.

Restricting \eqref{eq:whit} to be purely ingoing in the bulk fixes the ratio $\frac{C_2}{C_1}$.  The near-boundary behavior ($r \to 0$) of $\psi$ is then
\beq
\psi(r\rar0)\sim A\left(\omega,k\right)r^{\Delta_-}+B\left(\omega,k\right)r^{\Delta_+},
\eeq
where $\Delta_\pm = 1\pm\nu$ sets the scaling dimension of the dual operator $\Delta=\Delta_+$.
The retarded Green's function of the dual theory is then
\beq\label{eq:Gret}
G_R\left(\omega,k\right)=\frac{B\left(\omega,k\right)}{A\left(\omega,k\right)}= e^{-2\pi\imath\nu}\frac{\Gamma\left(-2\nu\right)\Gamma\left(\frac{1}{2}+\nu-\imath\frac{g}{2}\sqrt{\frac{\lambda}{\omega}}\right)}{\Gamma\left(2\nu\right)\Gamma\left(\frac{1}{2}-\nu-\imath\frac{g}{2}\sqrt{\frac{\lambda}{\omega}}\right)}\left(2\imath\sqrt{\lambda\omega}\right)^{2\nu},
\eeq
following \cite{MIT, MIT2}.  While there are apparent similarities between eq.~\eqref{eq:Gret} and the two-point functions calculated in \cite{MIT,MIT2}, we should note the fundamental differences that reflect the non-relativistic nature of the theory we are considering.

The $\nu$ that appears here is strictly real and positive, while for relativistic charged scalars, $\nu$ can take imaginary values as the gauge coupling increases.  The interpretation given by the authors in \cite{MIT} was that sufficiently strong electric fields induced pair production of the charged scalars.  Our results fit consistently within this interpretation as, for the non-relativistic theory, we do not see such behavior.  Since $\nu$ does not become imaginary with large $g$, we find that
\beq\label{eq:GretINF}
\lim_{g\rar\infty} G_R\left(\omega,k\right)=e^{-2\pi\imath\nu}\frac{\Gamma\left(-2\nu\right)}{\Gamma\left(2\nu\right)}\left(g\lambda\right)^{2\nu}.
\eeq
In particular we find that $G_R$ becomes independent of the frequency $\omega$.  We will further discuss this feature below. Note, as we turn off the background electric field $g\rar0$, eq.~\eqref{eq:Gret} becomes
\beq
\lim_{g\rar0} G_R\left(\omega,k\right)=e^{-2\pi\imath\nu}\frac{\Gamma\left(-2\nu\right)}{\Gamma\left(2\nu\right)}\left(\frac{\imath}{2}\sqrt{\lambda\omega}\right)^{2\nu}.
\eeq
This recovers the known power law behavior for the two-point function for non-relativistic scalars, \ie $G_R\left(\omega,k\right)\sim\omega^\nu$ \cite{KBJM}.

The prefactors of $\omega^\nu$ appearing in eq.~\eqref{eq:Gret} possess a non-trivial dependence on the frequency.  In particular, the combination $g \sqrt{\frac{\lambda}{\omega}}$ in the Gamma functions indicates the emergence of a new scale in the non-relativistic dual theory, which is not found in the relativistic theory considered in \cite{MIT, MIT2}.  This may be expected, as the gauge coupling $g$ scales as a velocity, which becomes a dimensionful quantity in the non-relativistic theory\footnote{In relativistic theories we are used to fixing units so that the two fundamental constants $\hbar=c=1$. In these natural units length and time are both measured in inverse eV, and so velocities are pure numbers. In the non-relativistic Schr\"odinger equation $c$ does not appear; instead one commonly fixes units so that $\hbar=m=1$, where $m$ is the particle mass. Since $E \sim m v^2$, this now implies that velocity squared divided by energy is a dimensionless combination, not velocity by itself.}.  It may be useful to think of $g$ as setting a velocity scale analogous to a Fermi velocity in the near-horizon limit.  Then the $\omega$-independence of \eqref{eq:GretINF} suggests that the response of the scalar field on the boundary becomes instantaneous in the large velocity limit.  Whether this analogy between $g$ and the presence of a Fermi surface can be made more precise is an open question.

\section{Conclusion}
In this work, we have explored the role that charged backgrounds play in the khronometric formulation of Ho\v rava gravity through static, spherically symmetric Einstein-\AE ther theory.  We have found that the causal boundaries appearing in Schwarzschild-type, both asymptotically flat and AdS, are present in the AdS-Reissner-N\"ordstrom background as well.  This provides yet more evidence that universal horizons are indeed a generic feature of the probe limit of khronometric Ho\v rava gravity.  An analytic solution for a charged black hole was also constructed in section 3.   However, we have found that only an asymptotically flat extremal Reissner-N\"ordstrom geometry solved the full equations of motion for the metric functions, \ae ther vector, and the scalar potential.

In exploring the near horizon $AdS_2\times R^2$ emergent geometry, we have calculated the two point retarded Green's function for non-relativistic charged scalars.  We have found that there is a striking similarity to the relativistic results that have been found previously, which may have been anticipated given the nature of the Ho\v rava solution we have found.  There are, however as of yet, unresolved issues with the results obtained in Section 4.  In the relativistic calculations, it was natural to regard the dual description of the near horizon geometry as a CFT emerging due to zooming in on a local section of a fractional Fermi surface.  It is unclear whether this language can be faithfully imported to the non-relativistic theory as we have an additional scale present, which could in this vernacular be interpreted as the Fermi velocity.  We will leave the resolution of this question for future exploration.

\acknowledgments

We would like to thank Michael Wagman for conversations during the early stages of this work.  This work was supported in part by by the U.S. Department of Energy under Grant No. DE-FG02-96ER40956.

\bibliography{hlrnbhbib}
\bibliographystyle{unsrt}

\end{document}